\documentclass[12pt]{iopart}

	\usepackage[dvips]{color}
	\usepackage{graphicx}

	\usepackage[l2tabu, orthodox]{nag} 

\usepackage{iopams}  
	\usepackage{bm} 

 \expandafter\let\csname equation*\endcsname\relax 
  \expandafter\let\csname endequation*\endcsname\relax 
	\usepackage{amsmath, amssymb}
	\usepackage{booktabs}
	\usepackage{dcolumn}

\usepackage[capitalize]{cleveref}  
	\def\mrm{\mathrm}
	
	\def\be{\begin{equation}}
	\def\ee{\end{equation}}
	
	\newcommand{\dd}[2]{\frac{\mrm{d} #1}{\mrm{d} #2}}
	\newcommand{\pdd}[2]{\frac{\partial {#1}}{\partial {#2}}}
	\newcommand{\comments}[1]{}

\begin{document}

\title{Coupling chemical networks to hydrogels controls oscillatory behavior}

\author{Daniel Reeves, Jorge Carballido-Landeira and Juan P\'{e}rez-Mercader}

\address{Department of Earth \& Planetary Sciences, Harvard University,
100 Edwin H. Land Boulevard, Cambridge, MA  02142-1204}
\ead{jcarballidolandeira@fas.harvard.edu}

\begin{abstract}
In this letter, we demonstrate that oscillations and excitable behavior can be imparted to a chemical network by coupling the network to an active hydrogel. We discuss two mechanisms by which the mechanical response of the gel to the embedded chemical reactant provides feedback into the chemistry. These feedback mechanisms can be applied to control existing chemical oscillations as well as create new oscillations under some conditions.  We analyze two model systems to demonstrate these two effects, respectively: a theoretical system that exhibits no excitability in the absence of a gel, and the Oregonator model of the Belousov-Zhabotinsky reaction in which the metal catalyst is intercalated into the polymer network. This work can aid in designing new materials that harness these feedbacks to create, control, and stabilize oscillatory and excitable chemical behavior in both oscillatory and non-oscillatory chemical networks.
\end{abstract}



\section{Introduction}
Nonlinear chemical networks are popular candidates for model systems of complex biological and biochemical phenomena. Such networks can harness a balance of positive and inhibitory feedback controls to generate oscillatory and excitable behavior in both time and space. The complex networks and idealized model systems serve for example as analogs for studying various biological processes such as clocks, embryo development, and nervous system behavior \cite{Tyson1980,Aulehla2004,Schnitzler2005,Fries2005,Dale2006,Fell2011}.

Recently, there is a growing interest in utilizing these networks to go beyond studying natural systems to generating new materials with smart and biomimetic properties \cite{Yoshida2002,Murase2009,Yoshida2010,Shiraki2012}.

For example, it has been shown \cite{Yoshida1996,Yoshida1999} that by anchoring the metal catalyst of the Belousov-Zhabotinsky reaction to the polymer gel network of a hydrogel, the oscillatory redox reaction induces swelling and deswelling of the gel.  This swelling can be utilized to induce biomimetic behaviors such as peristaltic motion \cite{Shiraki2012}.

These advances show that coupling a hydrogel to a chemical network induces new and useful behaviors in the gel. Herein we argue that the complementary effect is of equal interest \cite{Boissonade2003, Yashin2006Macromolecules, Boissonade2009, Yashin2012, Horvath2011}: the mechanical gel contributes additional feedback to the chemical network. The additional mechanical feedback controls the stability of chemical limit cycles, and in some cases, introduces new excitable and oscillatory behavior that is otherwise impossible without mechanical coupling. Such behavior has been predicted and demonstrated \cite{Boissonade2009,Horvath2011} using a bistable reaction driving a pH sensitive gel. In those experiments, the swelling and deswelling of the gel modulates the mixing rate between the gel pore-space and the well-mixed surrounding reservoir. The chemistry is thereby coupled to the gel via transport and mixing processes. 

The above work exploring transport induced chemical oscillations, as well as work on chemically coupled BZ-gels, serve as inspiration for the work in this letter. Here, we demonstrate the roles of mechanical feedback in two model chemical networks in which the chemistry is directly coupled to the polymer network itself. First, we analyze a simple theoretical network that exhibits excitable and oscillatory behavior {\em only} when mechanically {\em and} chemically coupled to the surrounding medium. Second, we show that stability and magnitude of preexisting characteristic oscillations can be controlled by varying the material parameters of the coupled hydrogel.

 We then discuss the several mechanisms by which the swelling and deswelling of the polymer network contribute feedback. By coupling a chemical network to an active hydrogel, the additional feedback mechanisms can be taken advantage of in designing systems with tightly controlled constant, excitable, and oscillatory behaviors.

\section{Mechanically coupled chemical network}\label{model}
\begin{figure}
	  \centering
	   \includegraphics*[width=3in]{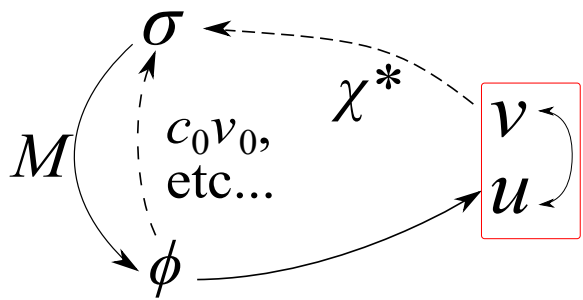}
		\caption{Schematic of chemomechanical and mechanochemical influence in a chemically-coupled hydrogel. The box indicates the bare chemical network. Solid and broken arrows indicate ``dynamic" and ``passive" influence, respectively, as described in the main text.}
\label{coupling diagram}
\end{figure}	

To understand the dynamical coupling described above, we consider a chemical reaction ongoing within the pore space of a hydrogel.  The reaction is characterized by two concentrations, $u$ and $v$, both measured with respect to the total volume (solvent and polymer), rather than solvent volume~\cite{Yashin2007}. The condition for this approximation is that the microscopic mixing lengthscale of polymer molecules is much smaller than the distance between reacting molecules. The chemical network is indicated in \Fref{coupling diagram} by the box and is characterized by two kinetic equations:
\begin{align}
\begin{split}
\pdd{u}{t} &= F(u,v) \\
\pdd{v}{t} &= G(u,v).
\end{split}
\label{eq: F and G general}
\end{align}

The chemical reaction is coupled to the enveloping hydrogel by embedding one of the reactants into the polymer network itself. For example, in BZ-coupled-hydrogels~\cite{Yoshida1996}, the metal catalyst is anchored via a bipyridine linker. As the concentration of coupled reactant changes the polymer network becomes more or less hydrophobic \cite{Yashin2006Macromolecules}. The internal stress, $\sigma$, therefore depends on the concentration of polymer-bound reactant, $v$.  In this way, the physical gel is influenced by the chemistry, as indicated by the arrow pointing from the box to $\sigma$ in \Fref{coupling diagram}.

Specifically, $\sigma$ depends on both $v$ and the polymer volume fraction, $\phi$. We model the dependence with the Flory-Huggins osmotic pressure for an isotropic elastic gel, modified to account for the interaction between solvent and anchored reactant\cite{Yashin2007}:
\begin{equation}
\frac{\sigma v_0}{KT} = -(\phi + \log(1-\phi) + \chi(T) \phi^2 ) + \chi^*v\phi + c_0 v_0 \left( \frac{\phi}{2\phi_0} - \left( \frac{\phi}{\phi_0}\right)^{1/3} \right),
\end{equation} 
where $\chi$ is the temperature dependent solvent-polymer coupling parameter, $c_0$ is the density of cross-linkers, $v_0$ is the volume of each statistical polymer subunit, $\phi_0$ is a reference polymer density, and $\chi^*$ is the parameter measuring the coupling between solvent and anchored reactant. The gel responds to changes in $\sigma$ by swelling and deswelling. Under the approximation that the gel is homogeneous in that reactants and polymer are evenly distributed throughout the gel, we can write
\begin{equation}
\dd{\phi}{t} = -M \phi^{1/6}(1-\phi) \sigma
\label{homogeneous phi}
\end{equation}
where $M$ measures the mechanical susceptibility, or mobility, of the gel. For further discussion of $\sigma$ and derivation of Eq. (3), we refer the reader to the Supplementary Information.

As the internal pressure varies, the gel expels and absorbs fluid thereby shrinking and swelling. This process contributes feedback from mechanics to the chemical kinetics in two ways.  First, non-limiting reactants with concentrations fixed relative to the outside solvent are diluted by the varying polymer fraction, $\phi$.  Suppose the concentration of such a reactant is $A$ in the bath.  When measured \emph{within} the gel relative to total volume, the concentration becomes $(1-\phi)A$~\cite{Yashin2007}. Therefore, each factor of $A$ in the kinetics is replaced by the $\phi$-dependent rate $(1-\phi)A$.

Second, the concentration of polymer increases as the gel contracts, causing an increase in the concentration of bound reactant, $v$. Conversely, as the gel swells, the concentration $v$ decreases.  On the other hand, the concentration of dissolved reactant, $u$, decreases as the gel contracts because the polymer behaves as a dilutant.  With this additional ``dilution'' effect, the time dependencies in Equation (1) become
\begin{align}
\begin{split}
\dd{u}{t} &= F(u(t),v(t),\phi(t)) - \frac{u}{1-\phi}\pdd{\phi}{t}   \\
\dd{v}{t} &= G(u(t),v(t),\phi(t)) + v \pdd{\phi}{t}.
\end{split}
\label{eq: PDEs}
\end{align}

\begin{figure}
	  \centering
	   \includegraphics*[width=4in]{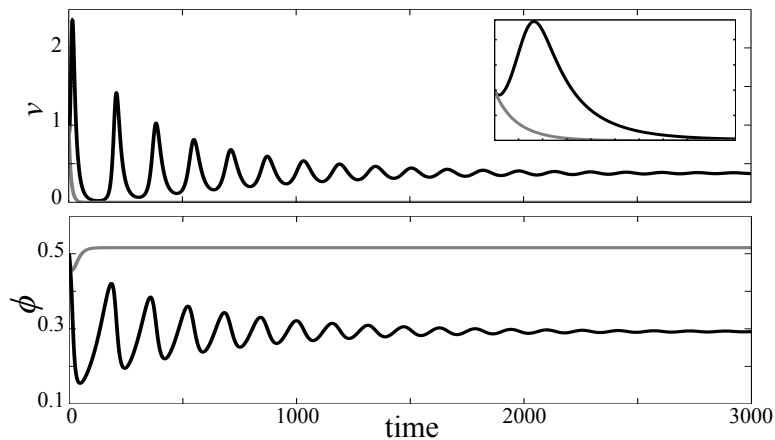}
		\caption{Damped oscillatory behavior demonstrated by the theoretical network. The top panel gives the concentration of $v$ both with the coupling (black line) and without (grey line). The former exhibits an initial jump whereas the latter rapidly decays to $v=0$, as shown in the detailed inset. The bottom panel gives the polymer density $\phi$. The model parameters are $\alpha=0.1$, $\beta=0.2$, $M=1$, $c_0v_0=3\times 10^{-3}$, $T=306$, and $\chi^*=0.045$.
}
	  	\label{figure: theoretical}
\end{figure}

In the homogeneous approximation diffusion is neglected as well as dilution of the dissolved species at the edges of the gel due to mixing of solvent with the reservoir. These effects are known to induce oscillations in some cases, as discussed in \cite{Yashin2007, Boissonade2009} and \cite{Horvath2011}.

In addition, we show that oscillations can be created and controlled in the absence of mixing processes. We apply the above coupled gel model to two chemical networks.  The first is a theoretical network of our design that exhibits no oscillatory or excitable behavior in the absence of an active gel:
\begin{align}
\begin{split}
F(u,v,\phi) &= 1-u(1-\phi)^2\\
G(u,v) &= \alpha u v - \beta v,
\end{split}
\label{eq: theory model}
\end{align}
where $\alpha$ and $\beta$ are dimensionless constants. However, we have introduced chemo-mechanical feedback by assuming that the degradation rate of the dissolved component, $u$, depends on non-limiting reactants and therefore inherits factors of $(1-\phi)$.  As we will see, this additional coupling introduces excitable behavior to the system.

The second model chemical network is a simplified form of the Oregonator, an approximate description of the Belousov-Zhabotinsky (BZ) reaction~\cite{Tyson1980}. For a thorough description of the BZ reaction and Oregonator, we direct the reader to \cite{Tyson1980,Krug1990}, and \cite{Amemiya2000}. In experimental studies, the chemical reaction is coupled to the hydrogel by linking the Ruthenium catalyst to the polymer \cite{Yoshida1996}. Accounting for fixed concentrations of nonlimiting reagents, the kinetics for the reactants are given by~\cite{Yashin2007}:
\begin{align}
\begin{split}
F(u,v,\phi) &= (1-\phi)^2 u -u^2 - f v(1-\phi) \frac{u-q(1-\phi)^2}{u+q(1-\phi)^2} \\
\epsilon G(u,v,\phi) &= (1-\phi)^2 u - (1-\phi)v,
\end{split}
\label{eq: FGP}
\end{align}
where $f$, $q$, and $\epsilon$ are well known BZ specific parameters.

The above dynamical systems are represented in \Fref{coupling diagram} which shows the direction of influence in the coupling among the chemistry (the box), the internal pressure $\sigma$, and the physical gel state $\phi$.  There are two classes of coupling between the mechanics of the gel and the chemistry. The first, shown in solid arrows, we refer to as ``dynamic" because dependencies are in the time derivatives of the variables.  The second, shown in broken arrows, we refer to as ``passive" because the dependencies are functional, but not explicitly in the time derivatives. The former are scaled by the mobility of the gel, $M$, whereas the latter are controlled by the polymer-catalyst coupling strength, $\chi^*$.  Thus, the model described above presents a chemo-mechanically and mechano-chemically coupled system.

\begin{table}
\caption{Parametric dependence of coupled Oregonator}
\resizebox{15.5cm}{!} {
\begin{tabular}{ll}
\toprule
Parameter &  \multicolumn{1}{c}{Behavior under increase} \\
\midrule
$\chi^*$ & Switches from small-type to large-type oscillations and generally increases amplitude\\
$c_0v_0$ &    Decreases amplitudes. Switches from large-type to small-type oscillations\\
$T$ &  Removes oscillations (Low $\chi^*$: decreases amplitudes; High $\chi^*$: increases amplitudes)\\
$M$ & Removes oscillations at low $M$, but introduces oscillations at high $M$\\
\bottomrule
\end{tabular}
}
\label{table: parameters}
\end{table}

\section{Results}\label{results}

\subsection{Theoretical network}
To show that coupling a non-oscillatory chemistry to a gel introduces new behaviors, we numerically integrate the theoretical model, Eqs. (4)-(5), for various values of $\alpha$ and $\beta$. We find excitable behavior for many parameter sets, including that shown in \Fref{figure: theoretical}, where the chemistry responds to a perturbation from the stationary state with an initial jump followed by damped oscillations. In the absence of coupling (where $(1-\phi)$ is constant), the chemical system quickly and smoothly returns to the steady state following a perturbation. Results of a linear stability analysis confirm these observations, that the uncoupled system will never oscillate and the coupled system can exhibit excitable jumps followed by damped oscillations. However, in the Supplementary Information, we present a similar three component model that exhibits undamped oscillatory behavior. Oscillatory behavior was previously reported for a non-oscillatory, but bistable, chemical network when immersed within a hydrogel \cite{Horvath2011}. In that system, the oscillatory behavior relies upon diffusive and advective communication between the bath and a bistable chemistry.  By coupling the reaction directly to the hydrogel, the requirements for both bistability and spatial resolution are therefore relaxed.

\subsection{Parametric control of oscillatory stability in Oregonator} 
In addition to creating new chemical oscillations, mechanochemical feedbacks modify the stability of existing oscillatory behavior.

Stationary states in the two-variable Oregonator are known to exhibit Hopf bifurcations under variations in the BZ parameters, leading to stable oscillations of $u$ and $v$. In the case of the BZ-coupled gel, linear stability analysis demonstrates that under the assumption of large mobility, $M$, variations in $\chi^*$ can destabilize the stationary state \cite{Yashin2006Macromolecules}.

We expand on that work by searching for stable oscillations while varying 4 parameters describing our coupled system which can be controlled experimentally via the design, condition, and preparation of the gel and reaction. The parameters are the mobility, $M$, solvent-catalyst coupling coefficient, $\chi^*$, crosslink density $c_0v_0$, and temperature $T$. For each set of parameters, we integrate Equation (4) from many initial conditions to find the magnitude of stable limit cycles. 
 The general behaviors observed by varying the four parameters are summarized in \Tref{table: parameters}.

\begin{figure}
	  \centering
	   \includegraphics*[width=4in]{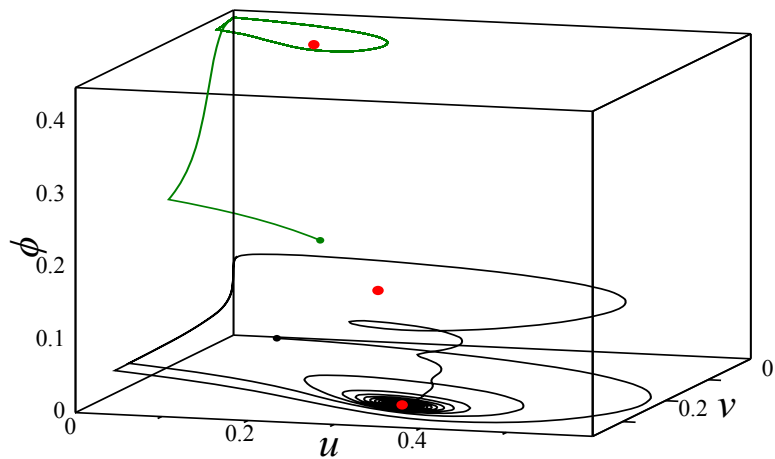}
		\caption{Trajectories of system in $u$, $v$ and $\phi$ originating at two distinct initial conditions (green and black dots) that approach stable oscillations near stationary states (red dots). The green trajectory forms a simple limit cycle around the deswollen fixed point whereas the black trajectory forms a ringing oscillation around the swollen fixed point. The parameters are $f=0.6$, $\epsilon=0.15$, $M=12.6$, $c_0v_0=3\times 10^{-4}$, $T=303$, and $\chi^*=0.0134$.}
	  	\label{two sample oscillations}
\end{figure}

As each parameter is increased, the stability and amplitude of oscillations change. These results are useful for designing BZ-coupled gels that exhibit engineered behaviors under varying conditions. More generally, the results demonstrate that the coupling causes the chemistry to adopt non-trivial and powerful dependencies on mechanical parameters. Therefore, the feedback mechanisms that are associated with these parameters offer new ways to control oscillatory chemical networks.

The addition of the mechanical variable, $\phi$, to the two-variable Oregonator permits new types of oscillatory behavior. If the mechanical dynamics vary over a timescale different from the chemical kinetics, ringing oscillations are observed as shown by the black curve in \Fref{two sample oscillations}. Each pulse consists of a slow compression of the gel followed by an excitation of the chemistry. The chemistry then rapidly oscillates around the $u-v$ stationary point as $\phi$ slowly decreases towards the global stationary point.

The ringing behavior depends on initial conditions; given one set of parameters, multiple stable oscillations are available and can be selected by initializing the system in the swollen or deswollen state, as shown by black and green curves in \Fref{two sample oscillations}.

\section{Discussion}\label{discussion}
\begin{figure}
	  \centering
	   \includegraphics*[width=4in]{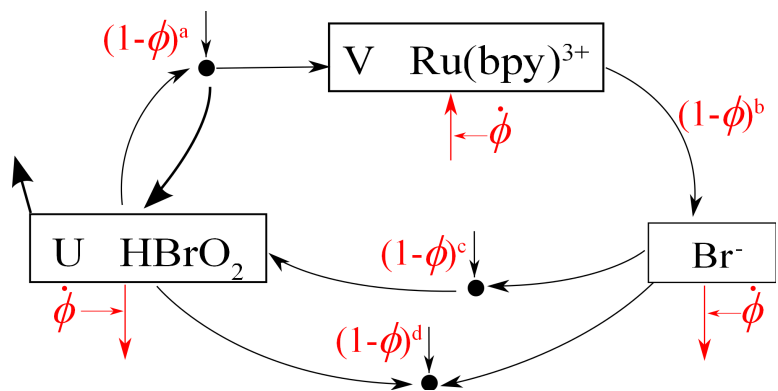}
		\caption{Two types of mechano-chemical feedback (in red) modifying the three-variable Oregonator (in black). The first type depends on $\dot{\phi}$ and the second depends on $(1-\phi)^n$. Reactants and influence lead into the reactions represented by black dots. The exponents appropriate for the BZ reaction are $a=2$, $b=d=1$, and $c=3$.
}
	  	\label{fig: feedback loop}
\end{figure}

Above, we describe how two novel feedback mechanisms are introduced by coupling gel and chemistry.
 The two mechanisms are illustrated in red in the context of the BZ reaction in \Fref{fig: feedback loop}. The species $\mrm{Br^-}$ is the inhibitor in the Oregonator, which is assumed to be at rapid equilibrium in our two-variable model. The first feedback mechanism is the modification of the kinetics due to the effective dilution of nonlimiting reactants by polymer. This gives rise to the $(1-\phi)^n$ terms. The second mechanism is the dynamic dilution and concentration of reactant concentrations due to the time derivative of $\phi$. This effect is notated in \Fref{fig: feedback loop} by arrows controlled by $\dot{\phi}$.

\begin{figure}
		\begin{center}
	  \centering
	   \includegraphics*[width=2.5in]{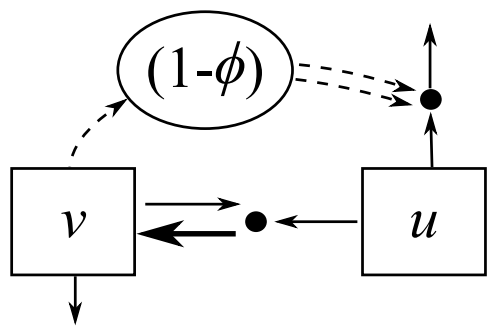}
\caption{Chemical schematic illustrating the role of $(1-\phi)$ in the theoretical network.}
	  	\label{figure: theory schematic}
			\end{center}
\end{figure}

The factors of $(1-\phi)^n$ influence the kinetics both dynamically and at steady state. From one perspective, the quantity $(1-\phi)$ acts as a ``reactant'' in its own right as reflected in Eq.(3). When $v$ increases, $(1-\phi)$ increases to approach osmotic equilibrium.  It is included in the system of differential equations almost on the same footing as $u$ and $v$. In this interpretation, the new ``reactant" acts as a bridge, as shown schematically in \Fref{figure: theory schematic}.

From another perspective, the dilution of the non-limiting reactants, as captured by factors of $(1-\phi)^n$, simply modify the kinetic rates. Therefore, the linear stability of the stationary states depends on the particular exponents. As the value $v$ is perturbed from the stationary state, the equilibrium value for $(1-\phi)$ is perturbed and the gel expands or contracts at a rate proportional to the mobility. The new value of $(1-\phi)$ feeds back to various reactions with a strength related to the exponent, whose value is determined by the stoichiometry. This feedback then either damps or magnifies the initial perturbation.

\begin{figure}
	  \centering
	   \includegraphics[width=6in]{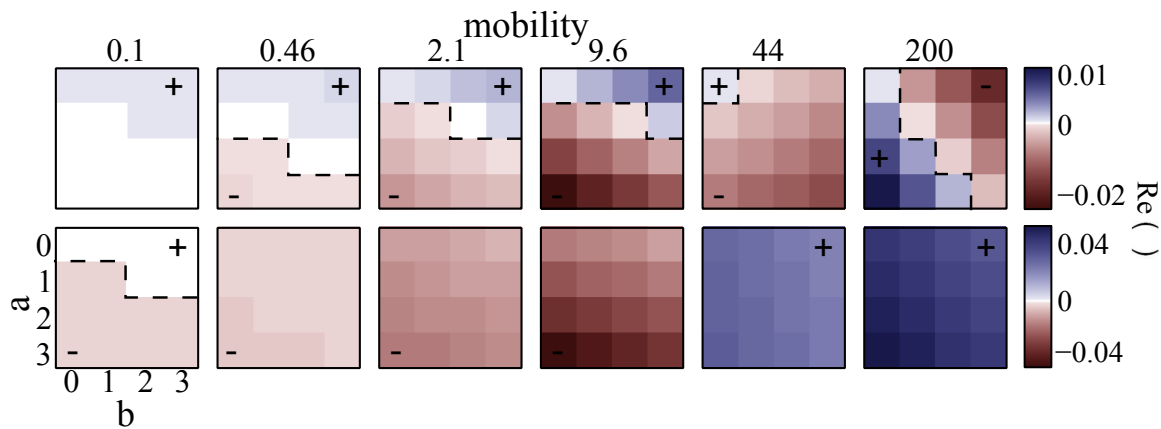}
		\caption{Stability of the system at stationary state for various mobilities and exponents $a$ and $b$.  Each square gives $a$ and $b = 0,1,2,3$ (for all cases $c=d=0$) for one value of the mobility. Blue indicates at least one eigenvalue has positive real part (unstable), whereas red indicates all negative real parts (stable). All factors of $(1-\phi)^n$ are normalized. The broken white line separates positive from negative. Top and bottom rows correspond to systems lacking and containing the $\dot{\phi}$ term, respectively.}
	  	\label{fig:evalues}
\end{figure}

We can quantify this effect by computing the largest real part of the eigenvalues of the linearized system at stationary state as a function of the exponents and the mobility, $M$ (see \Fref{fig:evalues} (top row)). To isolate the effect of feedback, we normalize factors of $(1-\phi)$ in Equation (4) by the stationary value, $(1-\phi_\mrm{SS})$. We find a nonmonotonic dependence on mobility where for small $M$ the stationary state is stabilized for $a>b$, but the opposite is true for large $M$. This rich variety of feedback mechanisms introduces nonintuitive effects as the mechanical mobility is adjusted.

We next examine \Fref{fig: feedback loop} to anticipate how the $\dot{\phi}$ terms influence the overall dynamics. Consider for example the rising side of a pulse, in which the gel is oxidizing.  As $v$ increases, the gel swells and $\dot{\phi}<0$.  The rate $\dot{v}$ consequently decreases. This contributes an ``inertia" to the system, in that it resists change in the oxidation state. Stationary states are unaffected, but oscillations, excitations, and transients will experience a ``drag".  This mechanism is responsible for the damped oscillations exhibited in \Fref{figure: theoretical}. The alternate model presented in the Supplementary Information swaps the sign of the drag term, thereby stabilizing the oscillations and avoiding the overdamped situation.  \Fref{fig:evalues} (bottom row) shows how the stability of the stationary state in the Oregonator is controlled by these inertial terms in a nonmonotonic way. In designing a chemical network, this drag effect can be utilized to resist rapid changes in the reactant concentrations, thus contributing an inductance-like effect in the chemical network.

\section{Conclusions}\label{conclusions}
In this work we have furthered the understanding of how chemomechanical and mechanochemical couplings control and induce chemical oscillations in coupled hydrogels. It has been suggested that such systems offer applications as biomimetic devices and micro-actuators \cite{Yoshida2010}.  The additional control described here could be critical in engineering finely tuned chemical oscillations in such devices as well as ensuring behavioral robustness. Our analysis helps identify perturbations that eliminate or induce oscillations.

More generally, we have enumerated and described the two feedback mechanisms that are general to hydrogel coupled chemical reactions: a multiplicative term and an inertial feedback. We have shown that these mechanisms can be harnessed to design systems that include new positive and negative feedback loops that introduce excitable and oscillatory dynamics to an otherwise simple non-oscillatory chemistry. We have shown that simple model systems gain control over oscillatory behavior by mechanochemically coupling them to the gel. In summary, hydrogel-coupled non-linear reactions can be employed to design and engineer complex networks utilizing a full toolbox of positive and negative feedbacks.

\ack
We thank Michael Stich for his useful comments regarding the manuscript.  We thank Repsol S.A. for funding this research. The funders had no role in study design, data collection and analysis, decision to publish, or preparation of the manuscript.

\section*{Supplementary Information}
\subsection*{Mechanical pressures}
Chemically cross-linked hydrogels are subject to internal forces arising from both elastic stresses and osmotic pressure. The latter is due to both enthalpic solvent-polymer interactions as well as the entropically unfavorable displacement of solvent molecules by polymer. 

We use the Flory-Huggins model of free energy in a polymer gel to derive the internal stress, $\sigma$, as a function of the polymer volume fraction, $\phi$. For an isotropically swollen gel, $\sigma$ is given by the sum of the osmotic pressure and the elastic stress \cite{Tomari1995}:
\begin{equation}
\frac{\sigma v_0}{KT}= -(\phi + \log(1-\phi) + \chi \phi^2 )+ c_0 v_0 \left( \frac{\phi}{2\phi_0} - \left( \frac{\phi}{\phi_0}\right)^{1/3} \right),
\end{equation} 
where $KT$ is the thermal energy, $c_0$ is the concentration of cross-linking molecules, $\phi_0$ is a reference polymer density, and $v_0$ is volume of one statistical subunit of gel at $\phi_0$.  The reference density is that at which the individual sub-units adopt Gaussian conformations with minimal elastic energy.   
$\chi$ is the polymer-solvent enthalpic interaction term. For a PNIPAAm gel Hirotsu \cite{Hirotsu1991} gives the enthalpic term as the sum of constant, temperature dependent, and density dependent parts:
\begin{equation}
\chi = \chi_0(T) + \chi_1 \phi
\end{equation} 
where $\chi_0 = 3.42 - 903/T$ and $\chi_1 = 0.518$.  

The overall enthalpic interaction is augmented by the additional interactions between solvent and reactive sidegroups anchored to the gel.  In the BZ-gels, these are solvent-catalyst and solvent-bipyridine interactions. We assume a general form of the interaction that varies with the oxidized fraction of catalyst, $f_\mrm{ox}$:
\begin{equation}
\chi = \chi_0(T) + \chi_1\phi + \chi_\mrm{bpy} n_\mrm{cat} + \chi_r n_\mrm{red} + \chi_\mrm{ox} n_\mrm{ox},
\label{implicit assumption}
\end{equation}
where $n_\mrm{cat}$ is the total number of catalyst molecules per statistical gel subunit, and $n_\mrm{ox}= f_\mrm{ox} n_\mrm{cat}$ and $n_\mrm{red}= (1-f_\mrm{ox})n_\mrm{cat}$ are the number of oxidized and reduced catalyst molecules, respectively. 

Under this assumed interaction, we can rewrite the interaction term as $\chi \phi^2 = \left(\chi_0(T) + n_\mrm{cat} (\chi_\mrm{red} + f_\mrm{ox} \Delta\chi_\mrm{ox}) + \chi_1 \phi \right)\phi^2$,
where $\chi_\mrm{red} = \chi_\mrm{bpy} + \chi_\mrm{r}$, and $\Delta \chi_\mrm{ox} = \chi_\mrm{ox}-\chi_\mrm{r}$.

We next express the oxidized fraction as $f_\mrm{ox} = v V_0 v_0/(n_\mrm{cat} \phi)$, where $v=V/V_0$ is the dimensionless concentration. 
 By substituting for $f_\mrm{ox}$, we rewrite the total stress as
\be
\frac{\sigma v_0}{KT} = -\phi - \log(1-\phi) - (\chi_0'+\chi_1\phi)\phi^2 - \chi^*v \phi + c_0 v_0 \left( \frac{\phi}{2\phi_0} - \left( \frac{\phi}{\phi_0}\right)^{1/3} \right),
\label{total stress}
\ee
where $\chi_0'= \chi_0(T) + n_\mrm{cat}\chi_\mrm{red} $ and $\chi^* = v_0 V_0 \Delta \chi_\mrm{ox}$.  We recover the effective form used in previous works \cite{Yashin2006Macromolecules, Yashin2006Science}, and expose Equation (9) as their underlying implicit assumption. Performing fits to published experimental data \cite{Yoshida1999}, we find approximate parameter values of $\chi_\mrm{red} = 12.5$ and $\Delta \chi_\mrm{ox} = -13.9$

\subsection*{Mechanical response}
The velocity at the edge of the gel is $w_p = \dd{r}{t}$ where $r$ is the radius of the deformed gel. This velocity is proportional to the gradient of the osmotic pressure \cite{Yashin2007}:
\begin{equation}
w_\mrm{p} = \lambda_0 \frac{\phi_0^{3/2}}{\phi^{3/2}} (1-\phi) \nabla \sigma,
\end{equation}
where $\lambda_0$ is a dimensionless constant describing the mobility of the gel \cite{Yashin2007}.  Assuming a homogeneous gel, we approximate the gradient of the pressure to be $\nabla \sigma \approx \sigma/r$:
\begin{equation}
\dd{r}{t} = \lambda_0 \frac{\phi_0^{3/2}}{\phi^{3/2}} (1-\phi) \frac{ \sigma}{ r}.
\end{equation}
For a spherical gel we compute $\mrm{d}\phi/\mrm{dt}$ by applying the relationship ${(r/R)^3 = (\phi_0/\phi)}$, where $R$ is the radius of the gel at $\phi_0$. Manipulation yields
\begin{equation}
\dd{\phi}{t} = -M \phi^{1/6}(1-\phi) \sigma,
\end{equation}
where $M=3 \lambda_0  \phi_0^{5/6}/R^2$.

\subsection*{BZ and Oregonator}
We model the BZ reaction utilizing a modified \cite{Yashin2007} and simplified \cite{Tyson1980} Oregonator.  The Oregonator captures two key feedback features of the BZ reaction. One feature is the autocatalytic production of the species $u$ via the transition of reduced catalyst into the oxidized catalyst, $v$. The second key feature is the inhibition of the reaction via the consumption of $u$ by a third reactant, $y$, which we assume to be in rapid equilibrium. 

The chemical kinetics are described by the two rates $\partial_t u = F(u,v,\phi)$ and $\partial_t v = \epsilon G(u,v,\phi)$, where $u$ and $v$ are reduced forms of $U$ and $V$, the concentrations of $\mrm{HBrO_2}$ and polymer-bound oxidized catalyst, respectively. The reduced concentrations are defined via $u = U/U_0$ and $v = V/V_0$, where $U_0$, $V_0$, the characteristic time, and the BZ parameters $\epsilon$, $f$, and $q$ are as defined in \cite{Yashin2007}.

The kinetic functions $F$ and $G$ are 
\begin{align}
\begin{split}
F(u,v,\phi) &=  u -u^2 - f v \frac{u-q}{u+q}, \\
G(u,v,\phi) &=  u - v.
\end{split}
\label{eq: FG}
\end{align}
where concentrations are written with respect to total volume, an approximation that is valid when the microscopic mixing lengthscale of polymer molecules is much smaller than the distance between reacting molecules. 

\begin{figure}
	  \centering
	   \includegraphics*[width=4in]{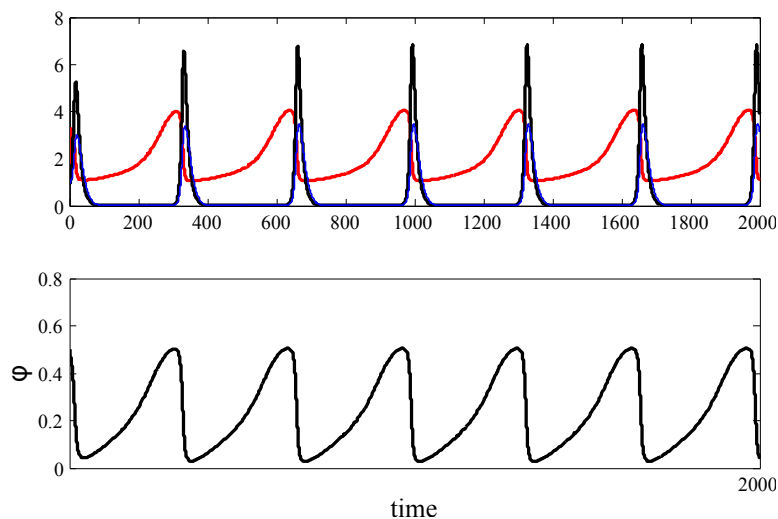}
		\caption{Integrated trajectory of theoretical model system.  The top panel gives the concentrations of $u$ (red), $v$ (black), and $z$ (blue) as a function of unitless time.  The lower panel gives the polymer volume fraction, $\phi$.  The parameters are $\alpha=0.1$, $\beta=0.2$, and $M=1$.
}
	  	\label{figure: theoretical 2}
\end{figure}

\subsection*{Oscillatory variant to theoretical network}
We introduce a variant of the theoretical network presented in the main text.  The two component model in the main text cannot sustain chemical oscillations but exhibits damped oscillations following an initial excitable jump in response to a perturbation from the stationary state.  This three component variant, however, exhibits stable oscillations. Here, both $u$ and $v$ are dissolved in solution and we introduce a new polymer-bound reactant, $z$.  The concentration of bound reactant $z$ simply tracks the concentration of $v$:
\begin{align}
\begin{split}
F(u,\phi) &= 1-u(1-\phi)^2\\
G(u,v) &= \alpha u v - \beta v\\
H(v,z) &= \beta(v-z)
\end{split}
\label{eq: theory model}
\end{align}
where $H(v,z)=\dot{z}$.  The pressure is a function of the new variable, $\sigma(z)$.  Since $z$ is enslaved to the auto-catalytic reactant $v$, increases in $v$ swell the gel. In response, the concentrations of the dissolved reactants $u$ and $v$ are increased.  In the two variable excitable model, the swelling that results from increases in $v$ simultaneously dilutes $v$, creating a drag on the system that damps the oscillations.  In this modified model, both $u$ and $v$ are concentrated upon swelling whereas $z$ is diluted.  This system undergoes stable oscillations as shown in \Fref{figure: theoretical 2}.

\section*{References}
\bibliography{mechanosensitivity}

\end{document}